\documentstyle[12pt]{article}

\begin{document}
\titlepage
\def\x{\times}
\def\ra{\rightarrow}
\def\beq{\begin{equation}}
\def\eeq{\end{equation}}
\def\beqa{\begin{eqnarray}}
\def\eeqa{\end{eqnarray}}
\def\D{ {\cal D}}
\def\L{ {\cal L}}
\def\C{ {\cal C}}
\def\calE{{\cal E}}
\def\lin{{\rm lin}}
\def\Tr{{\rm Tr}}
\def\mxth{\mathsurround=0pt }
\def\xversim#1#2{\lower2.pt\vbox{\baselineskip0pt \lineskip-.5pt
  \ialign{$\mxth#1\hfil##\hfil$\crcr#2\crcr\sim\crcr}}}
\def\simgr{\mathrel{\mathpalette\xversim >}}
\def\simle{\mathrel{\mathpalette\xversim <}}

\def\R{ {\cal R}}
\def\I{ {\cal I}}
\def\calR{ {\cal R}}
\def\lag{Lagrangian}
\def\Kahler{K\"{a}hler}

\begin{flushright} CERN-TH.6685/92  \\
UPR-0526T \end{flushright}

\vspace{4ex}

\begin{center}

{\bf NATURAL SUPERGRAVITY INFLATION }\\
         \rm
\vspace{3ex}
Gabriel Lopes Cardoso \\ and \\
Burt A. Ovrut\footnote{Work supported in part by the Department of Energy
under Contract No. DOE-AC02-76-ERO-3071. }\\
Theory Division, CERN, CH-1211 Geneva 23, Switzerland \\ and\\
Department of Physics, University of Pennsylvania,\\
Philadelphia, PA 19104-6396, USA \\

\vspace{3ex}

{\bf ABSTRACT}

\end{center}

We show that a single uncharged chiral superfield, canonically
coupled to \mbox{$N=1$} supergravity with vanishing
superpotential, naturally drives inflation in the early universe
for a class of simple Kahler potentials.  Inflation occurs due to
the one-loop generation of a Kahler anomaly proportional to
$\R^2$.  The coefficient of this $\R^2$ term is of the correct
magnitude to describe all aspects of an inflationary cosmology,
including sufficient amplitude perturbations and reheating.
Higher order terms proportional to $\R^n$ for $n \geq 3$ are
naturally suppressed relative to the $\R^2$ term and, hence, do
not destabilize the theory.  \\
\vspace{20 mm}
\begin{flushleft} CERN-TH.6685/92 \\
UPR-0526T \\
October 1992 \end{flushleft}

\newpage

It has been known for some time that one can achieve inflationary
growth in the early universe by modifying the usual Einstein
gravity \lag \cite{1}.  This is accomplished by adding the term
$\epsilon \calR^2$, where $\calR$ is the scalar curvature and
$\epsilon$ is a dimensionless constant.  Furthermore, it was
shown in \cite{2} that all the other physical requirements for an
inflationary cosmology, such as sufficient amplitude
perturbations and baryogenisis, are satisfied if parameter
$\epsilon$ lies in the range $10^{11} \simle \epsilon\simle7 \x 10^{15}$.
This scenario, although potentially a very simple theory of
inflation, suffers from two obvious deficiencies.  First, why
should one introduce an $\calR^2$ term and, second and more
seriously, why is $\epsilon$ so large when one would expect
it to be of order one?  These deficiencies were not addressed in
\cite{1,2}.  Furthermore, if one can introduce $\epsilon
\calR^2$, then what about terms like $\epsilon^{(n)}\calR^{n}$ where
$n \geq 3$?  Shouldn't these terms also be added and if not, why
not?  A careful study of the effect of such terms on inflation
was presented in \cite{3,4}.  It was shown     that
higher $\calR^n$ terms tend to destabilize the $\calR^2$
cosmological scenario if the $\epsilon^{(n)}$ coefficients are
sufficiently large.  It would appear then, that these
deficiencies would have to be overcome before the $\epsilon
\calR^2$ cosmology can be taken seriously.  In this paper, we
present a simple particle physics theory in which all of these
deficiencies are resolved in a natural manner.

Recently, it has become clear that the classical Kahler symmetry
of four-dimensional, $N=1$ supergravity-matter theories is, in
general, broken by an anomaly at the one-loop level.  Part of this
anomaly arises from the coupling of the composite Kahler
connection to two gauge fields via a triangle graph.  This
process leads to non-holomorphic corrections to gauge couplings
and threshold effects \cite{5,6}.  More importantly, from the
point of view of this paper, is another contribution to the
Kahler anomaly which arises from the coupling of the Kahler
connection to two gravitons via a triangle graph.  This process
was first introduced in \cite{5}.  For simplicity, this anomaly
was computed in \cite{5} with the gravitons on-shell, and was
found to be proportional to $(C_{mnpq})^2$ and $\calR_{mnpq}
\tilde{\calR}^{mnqp}$.  However, in this paper, we will show that
when this result is extended off-shell, a non-topological
term proportional to $\calR^2$ must arise due to the constraints
of supersymmetry.  That is, terms proportional to the square of
the scalar curvature arise naturally in $N=1$ supergravity
theories through the one-loop Kahler anomaly.  Furthermore, the
coefficient of $\calR^2$ is not a constant but rather is found to
be a specific function of scalar fields,  the Kahler potential.
We will show in this paper that, for a class of simple Kahler
potentials, this coefficient can be naturally very large, easily
in the range required in \cite{2}.  Finally, we find that all
radiative corrections involving $\calR^n$ for $n \geq 3$ are,
since they are Kahler invariant,           strongly suppressed
relative to the anomalous $\calR^2$ term.  Hence, these terms do
not destabilize the $\calR^2$ cosmology.  Thus a  class of  simple
four-dimensional, $N=1$ supergravity theories generically exhibit
$\epsilon \calR^2$ type cosmological behaviour, while naturally
resolving all the associated deficiencies.  The simplest such
theory, which we discuss in this paper, consists of a single
chiral superfield with vanishing superpotential coupled to
canonically normalized Einstein supergravity.  Remarkably, the
radiative corrections to this theory naturally lead to $\epsilon
\calR^2$ inflationary cosmology without any of the deficiencies.

We begin by presenting a way of generating $\calR^2$
terms with field dependent coefficients in the effective  action
for matter coupled to supergravity.  These $\calR^2$ terms arise
from one-loop triangle graphs that are anomalous under Kahler
transformations.  First, let us recall a few fundamental facts
about four-dimensional chiral matter coupled to supergravity.
The most general tree-level \lag\ describing the coupling of
chiral matter superfields $\Phi^i$ to supergravity is determined
by three functions \cite{a}; the Kahler potential $K(\Phi^i,
\Phi^{i\dag})$, the holomorphic superpotential $W(\Phi^i)$ and the
holomorphic gauge coupling function $f_{ab} (\Phi^i)$.  In this
paper, only the Kahler potential is needed for the
implementation of an inflationary phase in $\calR^2$ cosmology.
Hence, we will set $W(\Phi^i) = f_{ab} (\Phi^i) = 0$.  More
general
theories with $W(\Phi^i)  \neq 0$ and $f_{ab} (\Phi^i) \neq 0$
will be discussed elsewhere \cite{toapp}.  We note here, however,
that non-zero values for these functions need not, in general, alter
the conclusions of this paper.  The physical component fields of
a chiral matter superfield $\Phi^i$ are denoted by $\Phi^i \sim
(A^i, \chi^i)$.  The scalar component fields $A^i$ are
coordinates of a complex Kahler manifold with Kahler metric $g_{i
\bar{j}}$ \cite{b,c}.   Under Kahler transformations
\beq K(A^i, \bar{A}^i) \ra K (A^i, \bar{A}^i ) + F (A^i) +
\bar{F} (\bar{A}^i) \eeq 
where $F(A^i)$ is an arbitrary holomorphic function of $A^i$.
Simultaneously, all matter fermions $\chi^i$ rotate as \cite{a,c}
\beq \chi^i \ra e^{\frac{\kappa^2}{4} (F - \bar{F})} \chi^i \eeq 
where $\kappa^2 = 8 \pi$.
The relevant part   of the tree-level component field matter
\lag\ is given by
\beq \L = \ldots - i g_{i \bar{j}}\bar{\chi}^{\bar{j}}
\bar{\sigma}^m \D_m \chi^i + \ldots \eeq 
where
\beqa g_{i \bar{j}} &=& \partial_i \partial_{\bar{j}} K \nonumber
\\ \D_m \chi^i &=& (\partial_m - a_m) \chi^i + \ldots
\eeqa 
Invariance of this \lag\
under Kahler transformations is achieved
through the appearance of the Abelian Kahler connection $a_m$ in
the covariant derivative of $\chi^i$ \cite{c1}.  This connection is a
composite field given by
\beqa a_m &=& \frac{\kappa^2}{4} \left( \partial_i K \partial_m
A^i - \partial_{\bar{j}} K \partial_m \bar{A}^{\bar{j}} \right)
\nonumber \\
&& + \frac{i\kappa^2}{4} g_{i \bar{j}} \left( \chi^i \sigma_m
\bar{\chi}^{\bar{j}}\right) \eeqa 
Under Kahler transformations, it follows from (1) and (2) that $a_m$
transforms as an Abelian connection
\beq  a'_m =  a_m + \frac{\kappa^2}{4} \partial_m (F -
\bar{F}) \eeq 
and, hence, the \lag\ (3) is invariant.                    We
emphasize that any matter \lag\ coupled to supergravity, for any
Kahler potential $K$, automatically displays Kahler invariance at
the tree-level.  At the one-loop level, however, Kahler symmetry
might be anomalous.  In this paper, we are interested in the
subclass of triangle graphs  where two of the three external legs
are space-time spin connections $\omega_{ma}\,^b$.  The third
external leg is the Kahler connection $a_m$, while the matter
fermions $\chi^i$ run around the loop.
  This $a\omega \omega $ graph contributes a
complicated non-local expression to the one-loop effective action.
Its variation under Kahler transformations is, however,
readily obtained from the appropriate index-theorem \cite{d}.
The result is
\beq \delta_K S_{1- \rm loop} = - \frac{n\kappa^2}{768 \pi^2}
\int d^4 x
\sqrt{g} \I m F(A^i) \R_{mnb}\,^a \tilde{\R}^{mn}\,_a\,^b \eeq 
where $n$ denotes the number of matter superfields $\Phi^i$.
We now seek to find the supersymmetric extension of the component
expression (7).  Since at no stage has supersymmetry been
assumed to be broken, the $\calR\tilde{\calR}$ term in (7) should be
contained in the highest component of a superfield.  To find this
superfield, we will use some of the standard techniques employed
in deriving component field expressions from superspace
structures \cite{e}.  The bosonic curvature tensor $\R_{mnb}\,^a$
is related to the curvature superform            ${\rm R}_b\,^a =
\frac{1}{2!} dz^N dz^M {\rm  R}_{MNb}\,^a$ by
\beq \left. R_{mnb}\,^a \right| = \R_{mnb}\,^a \eeq 
where the bar denotes projection to the lowest component of the
superfield. We now use the fact that the
coefficients $R_{mnb}\,^a$ of the superform $R_b\,^a$ are
expressable \cite{f} in terms of the covariant supergravity
superfields $R$, $G_{\alpha \dot{\alpha}}$,
$W_{\alpha \beta \gamma}$ and their  hermitian
conjugates.  Then, using the decomposition properties of the
superform $R_b\,^a R_a\,^b$ \cite{e}, one readily finds that
\beq  \R_{mnb}\,^a \tilde{\R}^{mn}\,_a\,^b =
\left.  \frac{i}{16}  (\D^2 - 8  R^{\dag}) T \right| + h.c. + \ldots
\eeq 
where
\beq T = \left( \bar{\D}^2 - 8R\right) \left( 16 R^{\dag}R -
\frac{5}{2}
G_{\alpha \dot{\alpha}} G^{\alpha \dot{\alpha}} \right) + 32 W^{\alpha
\beta \gamma} W_{\alpha \beta \gamma} \eeq 
and where the deleted terms contain the gravitino and auxiliary
supergravity fields.
It then follows that the superfield expression for the anomaly
containing (7) is given by
\beq \delta_K S_{1 - \rm loop} = \frac{n}{4} \;
\frac{\kappa^2}{768
\pi^2} \int d^4 x d^2 \Theta \calE F(\Phi^i) T + h.c. \eeq
where $\calE$ is the chiral density superfield.
This equation can be readily integrated and yields, to the lowest
non-trivial order in supergravity fields,  the effective one-loop
\lag\ density
\beq \L_{\rm 1 - loop} = \frac{n}{64} \; \frac{\kappa^2}{768
\pi^2} \int d^2 \Theta \calE \bar{D}^2 \left[T\frac{1}{\Box} D^2
K (\Phi^i, \Phi^{i \dag})\right] + h.c. \eeq
Note that for on-shell gravitational
superfields, $R = G_{\alpha \dot{\alpha}} = 0$,
and hence expression (12) reduces to the
on-shell result given in \cite{5}. For a constant
classical background $\langle \Phi^i \rangle$, this \lag\
density  becomes
\beq \L_{1 - \rm loop} = \frac{n}{4} \; \frac{\kappa^2}{768
\pi^2} K \left( \langle \Phi^i \rangle, \; \langle \Phi^{i\dag}
\rangle \right) \int d^2 \Theta \calE    T + h.c. \eeq 
            We proceed to expand (13) out in components.  Using
the results in \cite{cs1}, we arrive at
\beqa \L_{1 - \rm loop} &=& \frac{n}{4} \; \frac{\kappa^2}{768
\pi^2} K \left( \langle A^i \rangle, \; \langle \bar{A}^i \rangle
\right) \left\{ - 2 \R^2 + \right. \nonumber \\
&&+ \left. 5 \left(\R_{mn} \right)^2 + 4 \left( \C_{mnpq}
\right)^2 \right\} + \ldots \eeqa 
where $\C_{mnpq}$ denotes the Weyl-tensor and the deleted terms contain
the gravitino and auxiliary fields.   Since we have assumed that the
$\langle A^i \rangle$ are all constant,     one can use
      the Gauss-Bonnet
theorem
\beq \int d^4 x \sqrt{g} \left\{\frac{2}{3}
 R^2-2 \left( R_{mn} \right)^2 + \left(
C_{mnpq}\right)^2                     \right\} =
   32 \pi^2 \chi \eeq  
where $\chi$ is the Euler number of the space-time manifold, to rewrite
expression (14)                  as
\beqa S_{1-\rm loop} &=& \frac{n}{4} \; \frac{\kappa^2}{768
\pi^2} \left[ \int d^4\right. x \sqrt{g} K \left( \langle A^i
\rangle, \; \langle \bar{A}^i \rangle \right) \left\{ -
\frac{1}{3} \R^2 + \frac{13}{2} \left( C_{mnpq} \right)^2
\right\} \nonumber \\ &&\left. - 80 \pi^2       K \left( \langle A^i
\rangle, \; \langle \bar{A}^i \rangle \right) \chi
\right] + \ldots
\eeqa 
We will drop the $\chi$ term, since it is a topological number.  We will
also drop the term proportional to the square of the Weyl tensor
$C_{mnpq}$ since it vanishes for a Robertson-Walker metric, which
is the type of space-time metric we consider in this paper.  This
one-loop action,      when added to the gravitational tree-level
action, yields the following effective gravitational \lag\
density for conformally flat space-time metrics
\beq \L_{\rm eff} = - \frac{1}{2 \kappa^2} \left\{ \R +
\frac{n}{72} K \left( \langle A^i \rangle, \; \langle \bar{A}^i
\rangle \right) \R^2 \right\} \eeq 
We have dropped the terms containing the auxiliary gravitational
fields, since they will not contribute  any $\R^2$-terms to (17).
We emphasize that expression (17) is only valid for constant $\langle
A^i \rangle$.  It will be shown below that this assumption is
justified
for the initial conditions relevant to inflation.
The effective \lag\ (17) is, obviously, not invariant under
Kahler transformations (1), since it includes the anomalous
diagram $a\omega \omega $.  A priori, there is nothing wrong with
a one-loop effective action which is not invariant under Kahler
transformations.  In fact, it provides a natural way of
generating an $\R^2$-term with a field dependent coefficient
which potentially can take any desirable value, small or large.
We point out that there is yet another anomalous triangle graph
capable of yielding such an $\R^2$-term, namely the triangle
graph $\Gamma\omega \omega$ where $\Gamma^i\,_{jk}$ is the
Christoffel connection of the complex scalar manifold with metric
$g_{i\bar{j}}$.  The graph $\Gamma\omega \omega $ is
anomalous under $\sigma$-model coordinate transformations of the
scalar fields $A^i$.  It contributes a term to the effective
action which is of the type (17) with $K$ replaced by $-2 \ln
\det g_{i \bar{j}}$ \cite{5,6}.  We will, however, ignore this
contribution, since it is negligible compared to (17) for the
choice of Kahler potential we use in this paper.  We close this
part of the discussion by writing down the field equations for
the effective gravitational \lag\ (17) coupled to matter fields.
 They are
\beqa \left( 1 + \frac{n}{36} K \R \right)
 \left( \R_{mn} - \frac{1}{2} g_{mn} \R \right) +
 \frac{n}{4(36)} K \R^2 g_{mn} + \nonumber\\
\frac{n}{36} K \left(\D_s \D_t \R \right)  \left( g_{mn}
g^{st} - \eta_m\,^s \eta_n\,^t \right)
= -\kappa^2 T_{mn}                 \eeqa  

We now discuss the classical evolution of a small isotropic and
homogeneous region of the universe, as determined by these field
equations.       The particular evolution we seek is divided into
three phases \cite{2}.  (i) The first phase is an inflationary
phase of superluminal expansion in which the Hubble parameter
decays linearly in time with a small slope. (ii) In the second
phase, the Hubble parameter approaches a zero value and bounces
back.  The universe goes into an oscillatory phase in which it is
reheated  as matter is excited by the oscillating geometry. (iii)
Finally, the universe goes over to a radiation dominated
         Robertson-Walker phase.  We begin by considering a small
isotropic and homogenous bubble with a      Robertson-Walker
metric
\beq ds^2 = -dt^2 + a^2 (t) \left[ dr^2 + r^2 \left( d \theta^2 +
\sin \theta d \phi^2 \right) \right] \eeq 
We take this region to be filled with a single chiral superfield
$\Phi \sim (A, \chi)$.  Furthermore, we choose the           Kahler
potential for $A$ to be
\beq K (A, \bar{A}) = e^{\bar{A} A} -1 \eeq 
Though unusual looking at first, this Kahler potential reduces to
the usual Wess-Zumino Kahler potential, $K = \bar{A} A$, for $|A|
\ll 1$.  Importantly, however, (20) differs dramatically from the
Wess-Zumino Kahler potential for $|A| > 1$.  Although we
will use (20) for concreteness on this paper, a large class of
simple Kahler potentials, such as $K= \frac{1}{e} \exp
(e^{\bar{A}A})-1$, would also lead to a satisfactory inflationary
cosmology.  The gravitational field equations were derived under
the assumption that $A$ is a constant, which we now denote as $A_i$.
In the homogenous bubble, we choose $A_i$ to be real and to lie
in the range $5.03 \simle A_i \simle 6.04$.  The crucial parameter which
governs the successful implementation of the inflationary
scenario we are about to discuss, is the function multiplying the
$\R^2$-term in (17), namely $\epsilon_i = \frac{1}{72} K$. The
above choice for the value $A_i$ translates into a value for
$\epsilon_i$ in the range $10^{11} \simle
\epsilon_i \simle 7 \x 10^{15}$.
This is precisely the range of values for $\epsilon_i$ shown
\cite{2} to be compatible with observational constraints on
scalar and tensorial perturbations (the upper bound) and the
requirements of standard baryogenesis and galaxy formation (the
lower bound).  The non-vacuous field equations are readily
obtained from (18) and (19).  The $t$-$t$ component of (18)
yields
\beq \dot{\R} = - \frac{1}{2} \frac{H}{\epsilon_i} - H \R +
\frac{1}{12} \frac{\R^2}{H} \eeq 
where $H= \frac{\dot{a}}{a}$ denotes the Hubble constant.
The curvature scalar $\R$ is given by
\beq \R = 6 \dot{H} + 12 H^2 \eeq 
Combining this equation and (21) yields
\beq \ddot{H} - \frac{1}{2} \frac{\dot{H}}{H} + 3 H \dot{H} +
\frac{1}{12} \; \frac{H}{\epsilon_i} = 0          \eeq
Next, we would like to solve (23) for $H$ in all three cosmological
phases
discussed above.  These solutions were first given in \cite{2}.
We discuss them briefly here, since the results will be used elsewhere
in this paper. We begin with the inflationary phase, defined by
$$ \left| \frac{1}{2} \; \frac{\dot{H}^2}{H} \right| \ll \left| 3 H
\dot{H} \right| \eqno{(24a)} $$
$$ \left| \ddot{H} \right| \ll \left| 3 H \dot{H} \right|
\eqno{(24b)} $$
\addtocounter{equation}{1}
Then, (23) becomes
\beq 3 H \dot{H} + \frac{1}{12} \frac{H}{\epsilon_i} = 0 \eeq 
which is solved by
\beq H = H_i - \frac{1}{36 \epsilon_i} t \eeq 
where we have set the time coordinate origin at $t_i = 0$.
Solution (26) describes a long phase in which $H$ decreases
linearly in time with a small slope.  It satisfies condition
(24b).
At the beginning of the inflationary phase, $t_i =0$, condition
(24a) becomes $\frac{1}{6\sqrt{6\epsilon_i}} \ll H_i$ and, hence,
imposes a lower bound on $H_i$.  Furthermore, it is clear from
(26) that condition (24a) will eventually be violated for large
$t$.  One can define the end of this inflationary phase to be the
time, $t_e$, when $\left| \frac{1}{2} \; \frac{\dot{H}^2}{H}
\right| = \frac{1}{10} \left| 3 H \dot{H} \right|$.  It follows
from (26) that
\beq t_e = 36 \epsilon_i \left( H_i - \frac{1}{6}
\sqrt{\frac{5}{3\epsilon_i}} \right) \eeq
Therefore, the inflationary period is defined by $0 \leq t \leq
t_e$.  In the oscillatory phase, on the other hand, all terms in
equation (23) are taken to be of comparable size.  Using the
ansatz
\beq H = f(t) \cos^2 \omega (t - t_0) \eeq
where $\omega$ and $t_0$ are constants to be specified below, and
by demanding that $\left( \frac{\dot{f}}{f}\right)^2 \ll
\frac{1}{6\epsilon_i}$, one finds
\beq f = \left( g_0 + \frac{3}{8\omega} \left[ 2 \omega \left[ t
- t_0 \right]  +  \sin \left( 2 \omega
\left[ t-t_0 \right] \right) \right] \right)^{-1} \eeq
where $\omega = \frac{1}{2 \sqrt{6 \epsilon_i}}$ and $g_0$ is an
arbitrary integration constant.  Parameter $t_0$ will correspond
to the beginning of the oscillatory phase if we further assume
that $\left( \frac{\dot{f}}{f} \right)^2 \left|_{t_0} =
\frac{1}{(16) 6 \epsilon_i}\right.$.  It follows from this
assumption that $g_0 = \frac{3}{\omega}$.  Other than the fact
that $t_e < t_0$, the value of $t_0$ has not yet been specified.
In the period defined by $t_e \leq t \leq t_0$, neither the
inflation solution (26) nor the oscillatory solution (29) is
strictly valid.  However, following \cite{2}, we will assume that
(26) is valid in this region and matches continuously onto (29)
at $t_0$.  It follows that
\beq t_0 = 36 \epsilon_i \left( H_i - \frac{1}{6}
\sqrt{\frac{1}{6\epsilon_i}} \right) \eeq
The oscillatory period continues until $t_F = t_0 + 12 \x 10^3
\frac{\epsilon_i^{3/2}}{N}$ where $N$ is the total number of matter
fields \cite{2}, at which time the Robertson-Walker phase begins.
Therefore, the oscillatory period is defined by $t_0 \leq t \leq
t_F$.  The only constant left unspecified thus far is the initial
Hubble parameter $H_i$.  This can be determined by integrating
equation (26) from $t_i = 0$ to $t_0$ to yield the total
expansion of the universe.  The result is
\beq a(t_0) = a_i e^{18 \epsilon_i H_i^2 - \frac{1}{12}} \eeq
To obtain approximately 60 $e$-followings of inflation, it
follows that
\beq H_i = \sqrt{\frac{10}{3\epsilon_i}} \eeq
which is larger than the lower bound discussed earlier.  It was
shown in \cite{2} that the above solutions lead to a satisfactory
theory of inflationary cosmology.  However, in our theory, it was
necessary to assume that $A$ is a spacetime constant in order to
obtain these results.  Is this assumption consistent?  To explore
this, let us take $A$ to be a real function of time but
independent of the spatial coordinates.  Then one can write
$A=A_i + A'(t)$ where $A_i$ is the parameter introduced above.
Furthermore, we assume that
\beq \left| \frac{A'}{A_i} \right| \ll \frac{1}{2A_i^2} \eeq
Under these conditions, the leading order equation of motion for
$A$ is relatively simple and found to be
\beqa \ddot{A}' + 3H \dot{A}' = - \frac{1}{72} \left( \frac{
A_i}{1 + A^2_i} \right) \left[ 3 \R^2 - 90 \left( \dot{H}^2 + 3
\dot{H}H^2 + 3 H^4 \right) \right] \eeqa
where $\R$ is given in (22).  Note that with the range of values
for $A_i$ discussed above, constant $A=A_i$ is only a solution of
(34) if $\R = H = 0$, which they are not.  It is necessary,
therefore, to go back to the original action (12) and to derive
the full Einstein equations allowing $A$ to be a function of
time.  We continue to assume that constraint equation (33) holds.
The result is a generalization of (18) which we need not present
here.  Suffice it to say that (18) is a very good leading order
approximation to the Einstein equations as long as
$$ \left| \dot{A}' A_i \right| \ll \frac{1}{6} \left|
\frac{\dot{\R}}{\R} \right| \eqno{(35a)}$$
$$ \left| (1+A^2_i) \dot{A}^{\prime 2} \right| \ll \frac{3}{8\pi}
H^2 \eqno{(35b)}$$
Therefore, our theory will closely approximate the constant
$A=A_i$ results, if the solutions of (23) satisfy (33), (35a) and
(35b) for all $t \geq 0$.
\addtocounter{equation}{1}
Let us assume that (33), (35a) and (35b) are satisfied. Then
equations (35) can be checked for consistency using the solutions
for $H$ given above in the inflationary and oscillatory periods.
In the inflationary period, where $0 \leq t \leq t_0$, $H$ is
given by (26).  If we further demand that $|\ddot{A}'| \ll |3
H\dot{A}'|$, then (34) simplifies to
\beq \dot{A}' = - \frac{1}{12} \left( \frac{A_i}{1+A_i^2} \right)
\left( \frac{\dot{H}^2 + 9 \dot{H} H^2 + 9 H^4}{H} \right) \eeq
At $t_i = 0$, it follows from (26) and (32) that
\beq \dot{A}'_i = - 5 \sqrt{\frac{5}{6}} \left( \frac{A_i}{1+A_i^2}
\right) \epsilon_i^{-3/2} \eeq
Furthermore, using this result, (22), (26) and (32), equations
(35a) and (35b) become $\epsilon_i \gg 10^3$ and $\epsilon_i \gg
10$ respectively, which are indeed satisfied for $\epsilon_i$ in
the range $10^{11} \simle \epsilon_i \simle 7 \x 10^{15}$.  Note that the
condition $|\ddot{A}'| \ll |3H\dot{A}'|$ becomes $\frac{1}{120}
\ll 1$ and, hence, is also satisfied.  Equation (33) is automatically
satisfied since $A'$ vanishes at $t_i = 0$.  It is tedious, but
straightforward, to show that (33), (35a) and (35b) remain valid
everywhere in the inflationary, oscillatory and Robinson-Walker
periods.  We conclude that our theory will closely approximate
the constant $A=A_i$ results for $\epsilon_i$ in the
cosmologically interesting range $10^{11} \simle
\epsilon_i \simle 7 \x 10^{15}$.

As discussed earlier, $\epsilon^{(n)} \R^n$ terms where $n \geq
3$ tend to destabilize the $\R^2$ cosmological scenario if the
$\epsilon^{(n)}$ coefficients are sufficiently large.  In this
paper, all such terms arise through radiative corrections.
Recall that our $\epsilon \R^2$ term was generated by a Kahler
anomalous, three-point one-loop graph involving one Kahler
connection and two spin connections.  Since Kahler symmetry is
violated, and since the anomaly must be non-local, the effective
\lag\ density was found to be of the form
\beq \L_{1 - \rm loop} \propto \kappa^2 \R^2 \frac{\Box}{\Box} K =
\kappa^2 \R^2 K \eeq
for constant $K$.  As we showed, $K$ (and, hence, $\epsilon$) can
be very large.  All other $\epsilon^{(n)} \R^n$ terms divides into
two types; a) those generated by graphs involving at least one
external Kahler connection and b) those generated by graphs all
of whose external legs are spin connections.  First consider type
a).  Since these graphs do not violate Kahler invariance, the
Kahler connection can only appear as the Kahler invariant field
strength $F_{Kmn} = \partial_m a_n - \partial_n a_m$.
Furthermore, these graphs must be local.  It follows that a graph
with $A$ Kahler connection legs and $B$ spin connection legs
generates an effective \lag\ density of the form
\beq \L'_{1- \rm loop} \propto (\kappa^2)^{A+B-2} \R^B F_K^A \eeq
For constant $K$, $F_K =0$ and, hence, $\L'_{1-\rm loop}$
vanishes.  Now consider type b).  These graphs generate an
effective \lag\ density of the form
\beq \L''_{1-\rm loop} \propto (\kappa^2)^{B-2} \R^B \eeq
These graphs do not, in general, vanish.  However, the
coefficients, not being field dependent, are always small numbers
of order unity.  Thus, since the relevant momenta are always
smaller than the Planck scale, these terms are much too small to
distabilize the cosmology generated by the $\epsilon \R^2$ term.

We close this paper with a brief, but important, discussion of the
input parameters in our theory.  As is well known, most
theories of inflationary cosmology suffer from the necessity
to introduce and fine-tune many input parameters.  Generically,
most theories, including ours, have five parameters which describe
conditions at the beginning of inflation.  In the notation of this
paper, these are $a_i$, $H_i$, $\R_i$, $A_i$ and $\dot{A}_i$.  In
our case, $a_i$ can have any value.  Furthermore, assuming (24)
and (35), which can be thought of as very mild tunings of order
$10^{-1}$ or $10^{-2}$, $\R_i$ and $\dot{A}_i$ can be expressed
in terms of $H_i$ and $A_i$ respectively.  Therefore, in our
theory, the only initial condition parameters are $H_i$ and
$A_i$.  In order to have at least 60 $e$-foldings of inflation,
it follows from (32) that $H_i \simle 10^{-5} - 10^{-8}$,
depending on  the value of $\epsilon_i$.  This is indeed a
fine-tuning, but it is considerably better than the value of
$H_i \simle 10^{-10}$ required in many other inflationary theories.
Finally, the range $5.03 \simle A_i \simle 6.04$ is of the order of
the Planck mass and, therefore, is not really a tuning at all.
Our theory shares this nice feature in common with chaotic inflation
scenarios \cite{16}, but is a vast improvement over most other
theories of inflation that must specify $A_i$ to a high  degree of
accuracy.  Hence, our initial condition parameters involve only
one, relatively mild, fine-tuning.  Furthermore, most other theories
of inflation must introduce a complicated scalar potential
energy with many new, highly fine-tuned parameters.  In our theory,
the input potential for $|A| \ll 1$
is almost irrelevant. We have chosen it to
be zero in this paper, but it can be non-vanishing as long as it
is naturally supressed for $|A| > 1$.  We believe, therefore, that
our claims of naturalness in the introduction are justified.

We would like to thank P. Steinhardt for pointing out to us the
relevance, and possible destabilizing effects, of the higher
$\R^n$ terms, and for other interesting conversations.

\newcommand{\PL}[3]{Phys. Lett. {\bf #1} (19#2) #3}
\newcommand{\NP}[3]{Nucl. Phys. {\bf #1} (19#2) #3}
\newcommand{\PR}[3]{Phys. Rev. {\bf #1} (19#2) #3}

\end{document}